\documentclass[aps,prl,twocolumn, superscriptaddress,showpacs,preprintnumbers,amsmath,amssymb]{revtex4}
\usepackage{color}
\usepackage{longtable}
\usepackage{graphicx} 
\usepackage{dcolumn}  

\graphicspath{{ps}}

\def\Journal#1#2#3#4{{#1} {\bf #2}, #3 (#4)}

\def\PLB{{ Phys. Lett.}  B}
\def\PRL{ Phys. Rev. Lett.}
\def\PRD{{ Phys. Rev.} D}

\def\chicx{\chi_{c{\rm J}}}
\def\NIMA{{ Nucl. Instrum. Methods Phys. Res., Sect. A }}

\begin{document}


\title{\quad\\[0.5cm] Evidence  of a new narrow resonance decaying to 
$\chi_{c1}\gamma$ in $B \to \chi_{c1} \gamma K$}

\noaffiliation
\affiliation{University of the Basque Country UPV/EHU, 48080 Bilbao}
\affiliation{University of Bonn, 53115 Bonn}
\affiliation{Budker Institute of Nuclear Physics SB RAS and Novosibirsk State University, Novosibirsk 630090}
\affiliation{Faculty of Mathematics and Physics, Charles University, 121 16 Prague}
\affiliation{University of Cincinnati, Cincinnati, Ohio 45221}
\affiliation{Deutsches Elektronen--Synchrotron, 22607 Hamburg}
\affiliation{Justus-Liebig-Universit\"at Gie\ss{}en, 35392 Gie\ss{}en}
\affiliation{Gifu University, Gifu 501-1193}
\affiliation{II. Physikalisches Institut, Georg-August-Universit\"at G\"ottingen, 37073 G\"ottingen}
\affiliation{Gyeongsang National University, Chinju 660-701}
\affiliation{Hanyang University, Seoul 133-791}
\affiliation{University of Hawaii, Honolulu, Hawaii 96822}
\affiliation{High Energy Accelerator Research Organization (KEK), Tsukuba 305-0801}
\affiliation{Ikerbasque, 48011 Bilbao}
\affiliation{Indian Institute of Technology Guwahati, Assam 781039}
\affiliation{Indian Institute of Technology Madras, Chennai 600036}
\affiliation{Institute of High Energy Physics, Chinese Academy of Sciences, Beijing 100049}
\affiliation{Institute of High Energy Physics, Vienna 1050}
\affiliation{Institute for High Energy Physics, Protvino 142281}
\affiliation{INFN - Sezione di Torino, 10125 Torino}
\affiliation{Institute for Theoretical and Experimental Physics, Moscow 117218}
\affiliation{J. Stefan Institute, 1000 Ljubljana}
\affiliation{Kanagawa University, Yokohama 221-8686}
\affiliation{Institut f\"ur Experimentelle Kernphysik, Karlsruher Institut f\"ur Technologie, 76131 Karlsruhe}
\affiliation{Korea Institute of Science and Technology Information, Daejeon 305-806}
\affiliation{Korea University, Seoul 136-713}
\affiliation{Kyungpook National University, Daegu 702-701}
\affiliation{\'Ecole Polytechnique F\'ed\'erale de Lausanne (EPFL), Lausanne 1015}
\affiliation{Faculty of Mathematics and Physics, University of Ljubljana, 1000 Ljubljana}
\affiliation{Luther College, Decorah, Iowa 52101}
\affiliation{University of Maribor, 2000 Maribor}
\affiliation{Max-Planck-Institut f\"ur Physik, 80805 M\"unchen}
\affiliation{School of Physics, University of Melbourne, Victoria 3010}
\affiliation{Moscow Physical Engineering Institute, Moscow 115409}
\affiliation{Moscow Institute of Physics and Technology, Moscow Region 141700}
\affiliation{Graduate School of Science, Nagoya University, Nagoya 464-8602}
\affiliation{Kobayashi-Maskawa Institute, Nagoya University, Nagoya 464-8602}
\affiliation{Nara Women's University, Nara 630-8506}
\affiliation{National Central University, Chung-li 32054}
\affiliation{National United University, Miao Li 36003}
\affiliation{Department of Physics, National Taiwan University, Taipei 10617}
\affiliation{H. Niewodniczanski Institute of Nuclear Physics, Krakow 31-342}
\affiliation{Nippon Dental University, Niigata 951-8580}
\affiliation{Niigata University, Niigata 950-2181}
\affiliation{Osaka City University, Osaka 558-8585}
\affiliation{Pacific Northwest National Laboratory, Richland, Washington 99352}
\affiliation{Panjab University, Chandigarh 160014}
\affiliation{University of Pittsburgh, Pittsburgh, Pennsylvania 15260}
\affiliation{Punjab Agricultural University, Ludhiana 141004}
\affiliation{RIKEN BNL Research Center, Upton, New York 11973}
\affiliation{University of Science and Technology of China, Hefei 230026}
\affiliation{Seoul National University, Seoul 151-742}
\affiliation{Soongsil University, Seoul 156-743}
\affiliation{Sungkyunkwan University, Suwon 440-746}
\affiliation{School of Physics, University of Sydney, NSW 2006}
\affiliation{Tata Institute of Fundamental Research, Mumbai 400005}
\affiliation{Excellence Cluster Universe, Technische Universit\"at M\"unchen, 85748 Garching}
\affiliation{Toho University, Funabashi 274-8510}
\affiliation{Tohoku Gakuin University, Tagajo 985-8537}
\affiliation{Tohoku University, Sendai 980-8578}
\affiliation{Department of Physics, University of Tokyo, Tokyo 113-0033}
\affiliation{Tokyo Institute of Technology, Tokyo 152-8550}
\affiliation{Tokyo Metropolitan University, Tokyo 192-0397}
\affiliation{Tokyo University of Agriculture and Technology, Tokyo 184-8588}
\affiliation{CNP, Virginia Polytechnic Institute and State University, Blacksburg, Virginia 24061}
\affiliation{Wayne State University, Detroit, Michigan 48202}
\affiliation{Yamagata University, Yamagata 990-8560}
\affiliation{Yonsei University, Seoul 120-749}
 \author{V.~Bhardwaj}\affiliation{Nara Women's University, Nara 630-8506} 
 \author{K.~Miyabayashi}\affiliation{Nara Women's University, Nara 630-8506} 
  \author{I.~Adachi}\affiliation{High Energy Accelerator Research Organization (KEK), Tsukuba 305-0801} 
  \author{H.~Aihara}\affiliation{Department of Physics, University of Tokyo, Tokyo 113-0033} 
  \author{D.~M.~Asner}\affiliation{Pacific Northwest National Laboratory, Richland, Washington 99352} 
  \author{V.~Aulchenko}\affiliation{Budker Institute of Nuclear Physics SB RAS and Novosibirsk State University, Novosibirsk 630090} 
  \author{T.~Aushev}\affiliation{Institute for Theoretical and Experimental Physics, Moscow 117218} 
  \author{T.~Aziz}\affiliation{Tata Institute of Fundamental Research, Mumbai 400005} 
  \author{A.~M.~Bakich}\affiliation{School of Physics, University of Sydney, NSW 2006} 
  \author{A.~Bala}\affiliation{Panjab University, Chandigarh 160014} 

% ...
\collaboration{The Belle Collaboration}


\begin{abstract}
We report measurements of $B \to \chi_{c1} \gamma K$ and 
$\chi_{c2} \gamma K$
decays using $772 \times 10^{6}$ $B\overline{B}$ events collected at
the $\Upsilon(4S)$ resonance with the Belle detector at the KEKB 
asymmetric-energy $e^+e^-$ collider. Evidence of a new resonance 
in the $\chi_{c1} \gamma$ final state is found 
 with a statistical significance of $3.8~\sigma$. 
This state has a mass of
$3823.1 \pm 1.8 \mbox{(stat)} \pm 0.7 \mbox{(syst)}$ MeV/$c^2$, a value that is
consistent with theoretical expectations for the 
 previously unseen $1^3 D_2$ $c\bar{c}$
meson. 
We find no other narrow resonance and set upper limits on
 the branching fractions of the $X(3872) \to \chi_{c1} \gamma$ and 
$\chi_{c2} \gamma$ decays.

\end{abstract}

\maketitle

During the last decade, a number of new charmonium 
($c\bar{c}$)-like states were observed, many of which are candidates for 
exotic states~\cite{brambilla}. The first of these,
the $X(3872)$,  has 
been observed  by six different experiments  in the same final 
state~\cite{belle1,cdf1,do1,babar1,lhcb,cms}. A recent update from 
Belle~\cite{belle_recent} and LHCb~\cite{lhcb} results
 in  a world average 
mass at $3871.68\pm 0.17$  MeV$/c^2$~\cite{pdg} and a stringent upper 
bound on its width ($\Gamma < 1.2$ MeV)~\cite{belle_recent}. The proximity 
of its mass  to the $D^{*0}\bar{D^0}$ threshold 
makes it a good candidate for a $D\bar{D}^*$ molecule~\cite{swanson}.
Other alternative models have been proposed, such as 
a tetraquark~\cite{Lmaiani}  or a $c\overline{c}g$ hybrid meson~\cite{Lihybrid}.


Radiative decays  can illuminate clearly
the nature of hadrons. For example, the observation of 
$X(3872) \to J/\psi \gamma$  confirmed  the $C$-even 
parity assignment for the $X(3872)$
\cite{babarprl102,belle3}.  
 The $X(3872)\to \chi_{c1} \gamma$ and $\chi_{c2}\gamma$ decays are
forbidden by $C$-parity conservation in 
electromagnetic processes. However, if
the $X(3872)$ is a tetraquark  or a molecular state,
it  may have a $C$-odd partner, which  could decay  into $\chi_{c1}\gamma$ 
and  $\chi_{c2}\gamma$ final states~\cite{terasaki,nieves}.

 In the charmonium family, the observation of a 
$D$-wave $c\bar{c}$ meson and its decay modes would test phenomenological 
models~\cite{cornell,buchmuller}.  The as-yet  undiscovered 
$1^3D_2~c\bar{c}$ $(\psi_2$) and $1^3D_3~c\bar{c}~(\psi_3)$ states are
expected to have significant branching fractions to $\chi_{c1}\gamma$ and 
$\chi_{c2}\gamma$, respectively~\cite{estia2002,cho1994}.  
$D$-wave $c\bar{c}$ states and their properties were predicted 
 long ago but  remain unconfirmed~\cite{estia2002,cho1994}. 
The E705 experiment reported an indication of a $1^3D_2$ state in
$\pi^{\pm}N\to J/\psi \pi^+ \pi^- + $
anything, with a mass of $3836\pm 13$ MeV$/c^2$~\cite{e705}; however, the 
statistical significance of this result was below the threshold
for  evidence.


In this letter, we report measurements of $B \to \chi_{c1} \gamma K$ and 
$B \to \chi_{c2} \gamma K$ decays, where  the $\chi_{c1}$ and $\chi_{c2}$ 
decay to $J/\psi \gamma$~\cite{mixchg}. 
These results are obtained from a data sample of 
$772\times 10^{6}$ $B\overline{B}$  events collected with the Belle 
detector~\cite{abashian} at the KEKB asymmetric-energy 
$e^+e^-$ collider operating at the $\Upsilon(4S)$ resonance~\cite{kurokawa}.


The $J/\psi$ meson is reconstructed via its decays to $\ell^+\ell^-$ 
($\ell =$ $e$ or $\mu$).  To reduce the radiative tail in
the $e^+e^-$ mode, the four-momenta of all 
photons within 50 mrad with respect to the original direction
of the $e^+$ or $e^-$ tracks are 
included in the invariant mass calculation, 
hereinafter denoted as  $M_{e^+e^- (\gamma)}$. 
The reconstructed invariant mass of the $J/\psi$ candidates is required 
to satisfy 2.95 GeV$/c^2 < M_{e^+ e^-(\gamma)} < 3.13$ GeV$/c^2$ or 
3.03 GeV$/c^2 < M_{\mu^+ \mu^-} < 3.13$ GeV$/c^2$. 
For the selected $J/\psi$ candidates,  a vertex-constrained fit is
applied and then a  mass-constrained fit is performed in order to improve 
the momentum resolution.   
The $\chi_{c1}$ and $\chi_{c2}$ candidates are reconstructed by combining
 $J/\psi$ candidates with a photon having energy ($E_{\gamma}$)  
larger than 200 MeV in the laboratory frame.  
Photons are reconstructed from energy depositions in  the
electromagnetic calorimeter (ECL), which do not match any
extrapolated charged track. 
To reduce the 
background from $\pi^0 \to \gamma\gamma$, we use a likelihood function
that distinguishes an isolated photon from $\pi^0$ decays using the photon pair
invariant mass, photon laboratory energy and polar angle~\cite{kopenberg}.  
We reject both $\gamma$'s in the pair if the 
$\pi^0$ likelihood probability is  larger than 0.7.
The reconstructed invariant 
mass of the $\chi_{c1}$  ($\chi_{c2}$) is required to satisfy 
3.467 GeV$/c^2 < M_{J/\psi\gamma} <$ 3.535 GeV$/c^2$ 
(3.535 GeV$/c^2 < M_{J/\psi \gamma} <$ 3.611 GeV$/c^2$). 
 A mass-constrained fit is applied to the selected 
$\chi_{c1}$ and $\chi_{c2}$ candidates.


Charged kaons are identified by combining information from 
the central drift chamber,  time-of-flight scintillation counters, 
and the aerogel Cherenkov counter systems. The kaon identification efficiency 
is $89\%$ while the probability of misidentifying a pion as a kaon  
is $10\%$. $K^0_S$ mesons are reconstructed by combining two oppositely charged 
pions with  the invariant mass $M_{\pi^+\pi^-}$ lying between 
482 MeV$/c^2$ and 514 MeV$/c^2$. The selected candidates are 
required to  satisfy the  quality criteria 
described in Ref.~\cite{goodks}.


To reconstruct $B$ candidates, each $\chicx$~\cite{CX} is combined 
with a kaon candidate and a photon having $E_{\gamma} > $ 100 MeV 
(and not used in the reconstruction of $\chicx$). 
If the invariant mass
of  any photon pair that includes this photon
is found to be consistent with a $\pi^0$ (i.e., 117 MeV$/c^2 < M_{\gamma\gamma} <$  
153 MeV$/c^2$), this photon is rejected.
 Among the events containing at least one $\chicx$ candidate,
9.0\% have multiple $\chicx$ candidates. In such cases, the $\gamma$
forming the $\chicx$ candidate with mass closest to the
$\chi_{c1}$ or
$\chi_{c2}$ masses~\cite{pdg} is not used as the additional photon. This 
treatment suppresses reflections from the $\chicx$ daughter photons.


The $B$ candidate is identified by two kinematic variables: 
the beam-constrained mass 
($M_{\rm bc} \equiv \sqrt{{{E}^{*2}_{\rm beam}} - {p^{*2}_{B}}}$) and the energy 
difference ($\Delta E \equiv E_{B}^*- E^*_{\rm beam}$). 
Here,  ${E^*_{ \rm beam}}$ is 
the run-dependent beam energy, and ${E^*_{B}}$ and ${p^*_{B}}$  are the 
reconstructed energy and momentum, respectively, of the $B$ meson candidates 
in the center-of-mass (CM) frame. Candidates within a $\Delta E$ window of 
$[-28, 30]$ MeV and with $M_{\rm bc} >$ 5.23 GeV$/c^2$ 
are  selected. Of these, 9.8\% (6.4\%) have multiple 
candidates in the $B^\pm \to\chi_{c1} \gamma K^\pm$ 
($B^\pm \to \chi_{c2} \gamma K^\pm$) mode; 
we select the $B$ candidate with $\Delta E$ closest to zero.
In order to improve the resolution in $M_{\chicx \gamma}$,
we scale the energy of the $\gamma$  so that $\Delta E$ is equal to zero. 
 This corrects for incomplete energy measurement in the ECL.
To suppress continuum background, events having a ratio $R_2$ of the second to 
zeroth Fox-Wolfram moments~\cite{foxwolfram} above $0.5$ are rejected.

\begin{figure}[h!]
\begin{center}
\includegraphics[height=70mm,width=87mm]{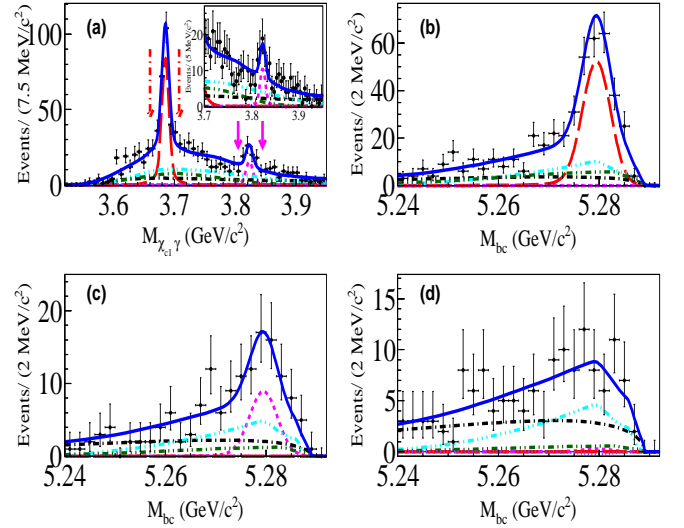}
\caption{\label{fig:c1gk} 2D UML fit projection for $B^{\pm} \to (\chi_{c1} \gamma) K^{\pm}$ decays:
(a) $M_{\chi_{c1}\gamma}$ distribution for $M_{\rm bc} > 5.27 $ GeV$/c^2$,
(b) $M_{\rm bc}$ distribution for 
3.660 GeV$/c^2 < M_{\chi_{c1}\gamma} < 3.708$ GeV$/c^2$ ($\psi'$ region,  
shown by   the red dotted-dashed arrows  in a),
(c) $M_{\rm bc}$ distribution for  
3.805 GeV$/c^2 < M_{\chi_{c1}\gamma} < 3.845$ GeV$/c^2$ ($X(3823)$ region,  
shown by  the magenta solid arrows in a) and
(d) $M_{\rm bc}$ distribution for  
3.84  GeV$/c^2 < M_{\chi_{c1}\gamma} < 3.89$ GeV$/c^2$ ($X(3872)$ region). 
The curves used  in the fits are described in~\cite{CURVES}.
 }
\end{center}
\end{figure}

The $M_{\chi_{c1}\gamma}$ and $M_{\rm bc}$ projections for 
the $B^{\pm} \to \chi_{c1} \gamma K^{\pm}$ signal candidates
are shown in Fig.~\ref{fig:c1gk}, where a $\psi' \to \chi_{c1} \gamma$ 
signal is evident. In addition, there is a significant narrow peak at
3823 MeV/$c^2$, denoted   hereinafter as $X(3823)$. No signal
of $X(3872) \to \chi_{c1}\gamma$ is seen.
We extract the signal yield from a two-dimensional unbinned extended 
maximum likelihood (2D UML) fit to the variables $M_{\chicx \gamma}$ and 
$M_{\rm bc}$. 


The resolution in $M_{\chicx\gamma}$ ($M_{\rm bc}$) is parameterized by a sum of 
two Gaussians (Gaussian and logarithmic Gaussian~\cite{lg}).
MC studies show that the resolutions in both 
$M_{\chicx \gamma}$ and $M_{\rm bc}$ for a narrow resonance in the mass range
3.8 GeV$/c^2 < M_{\chicx \gamma} <$  4.0 GeV$/c^2$  are in good 
agreement with those for $\psi'$. 
 The parameters of the resolution functions are determined 
from the MC simulation that is 
calibrated using the $B^\pm \to\psi'(\to\chi_{c1}\gamma)K^\pm $ signal.
We take into account the $\psi'$ natural width~\cite{pdg} by convolving 
the Breit-Wigner function and the resolution function;  for the $X(3823)$ 
and $X(3872)$, zero natural widths are assumed. The two-dimensional 
probability density function (PDF) is a 
product of the one-dimensional distributions. 


For $B^\pm \to \psi'(\to \chi_{c1}\gamma) K^\pm$ decays, the  mean and width of 
the core Gaussian are floated and the remaining parameters are fixed 
according to MC simulations. To fit  the $B^{\pm} \to X(3823) (\to \chi_{c1}\gamma) K^{\pm}$
signal, we float the mean of the core Gaussian  but constrain the detector 
resolution  by using the $\psi'$ signal results after taking 
into account the difference estimated from the signal MC study. 
For $M_{\rm bc}$, the parameters are fixed to those found for the $\psi'$, in 
accordance with expectations based on the MC simulation. 
For $X(3872)$, we fix the mass difference and the mass resolution change 
with respect to $\psi'$ using the  information from PDG~\cite{pdg} and MC studies.
To fit $B^0 \to \psi' (\to \chi_{c1}\gamma) K^0$,
$B^0 \to X(3823) (\to \chi_{c1}\gamma) K^0$,
$B \to \psi' (\to \chi_{c2}\gamma) K$,  
$B \to X(3823)(\to \chi_{c2}\gamma) K$ and 
$B \to X(3872)(\to \chicx \gamma) K$, we fix all the parameters obtained from 
the signal MC study after correcting  the PDF shapes by applying MC/data 
calibration factors.


To study background with a real $J/\psi$, we use large MC
simulated $B \to J/\psi X$ samples corresponding to 100 times the 
integrated luminosity of the 
data. The non-$J/\psi$ (non-$\chicx$) background is studied 
using $M_{\ell\ell}$ ($M_{J/\psi\gamma}$) sidebands in data. 
In $B \to (\chicx \gamma) K$, 
the background with a broad peaking structure is mostly due to the
$B \to \psi'(\nrightarrow \chicx \gamma) K$, $B \to \chicx K^*$,
$B \to J/\psi K^*$ and $B \to \psi' K^*$ decay modes. 
$B \to \psi' (\nrightarrow \chicx \gamma) K$
produces peaks in both distributions 
($M_{\chicx \gamma}$ and $M_{\rm bc}$), while the other
backgrounds are flat in 
$M_{\chicx \gamma}$ but  peaked in  $M_{\rm bc}$. We  determine the PDFs
from the large MC sample. The fractions of the PDF components are floated 
in the fit, except for $B \to \psi'(\nrightarrow \chicx \gamma) K$,
whose fraction 
is controlled by fixing its ratio to the $B \to \psi'(\to \chicx \gamma) K$ 
signal yield. For the combinatorial background, a threshold function, 
$(M_{\chicx \gamma})^2 \times  \exp(a ~(M_{\chicx \gamma} - M_{\rm th}) 
~+~b~(M_{\chicx \gamma} - M_{\rm th})^2~+~c~(M_{\chicx \gamma} - M_{\rm th})^3)$, 
where $M_{\rm th} = 3.543$ GeV$/c^2$ (3.585 GeV$/c^2$) for $M_{\chi_{c1}\gamma}$ 
($M_{\chi_{c2}\gamma}$), is used for $M_{\chicx \gamma}$  and an 
ARGUS function~\cite{argus} is used for 
$M_{\rm bc}$. The value of  $M_{\rm th}$ is estimated from 
a MC study; its variation, which affects the signal yield in the
fits, is incorporated in the systematic errors.
The $\Delta E$ data sidebands are used to verify  the background PDFs.
The fractions for the signal and the background components
 are floated in the fit.

The results of the fits are presented in Figs.~1-3
and in Table~\ref{tab_results}.
The significance is estimated using  the  value of
$-2\ln(\mathcal{L}_{0}/\mathcal{L}_{\rm max})$ where 
$\mathcal{L_{\rm max}}$ ($\mathcal{L_{\rm 0}}$) denotes the likelihood value 
when the yield is allowed to vary (is set to zero). 
In the likelihood calculation, the $\chi^2$ statistic uses the
 appropriate number of degrees of freedom
(two in the case of $B^\pm \to X(3823) K^\pm$ and one for  the other  decay
 modes). The systematic uncertainty, which is described below, 
is included in  the significance calculation~\cite{cousinhighland}. We find a 
significant $\psi'$ signal in all considered channels.
 We also obtain evidence for the $X(3823)$ in the channel  
$B^\pm \to \chi_{c1} \gamma K^{\pm}$ with a statistical significance of 3.8 standard deviations ($\sigma$). 
The $X(3872)$ signals are insignificant. We estimate the branching 
fractions according to the formula
$\mathcal{B} = Y/(\epsilon ~  \mathcal{B}_s ~ N_{B\bar{B}})$; 
here, $Y$ is the yield, $\epsilon$ is the reconstruction efficiency, 
$\mathcal{B}_s$ is the secondary branching fraction taken from 
Ref.~\cite{pdg} and $N_{B\bar{B}}$ is the number of $B \bar{B}$ mesons 
in the data sample. Equal production of neutral and 
charged $B$ meson pairs in the $\Upsilon(4S)$ decay is assumed.
Measured branching fractions for the $\psi'$ are in agreement with the world 
average values for all the channels~\cite{pdg}. 
We set  90\% confidence level (C.L.) upper limits (U.L.)  on the insignificant 
channels using frequentist methods based on  an ensemble of pseudo-experiments.



\begin{figure}[h!]
\begin{center}
\includegraphics[height=70mm,width=87mm]{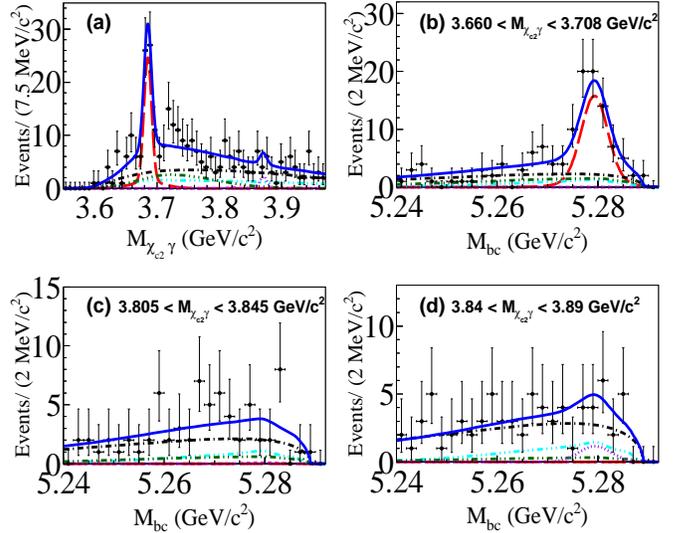}
\caption{\label{fig:c2gk} 2D UML fit projection for $B^{\pm} \to (\chi_{c2} \gamma) K^{\pm}$ decays:
(a) $M_{\chi_{c2}\gamma}$ distribution for $M_{\rm bc} > 5.27 $ GeV$/c^2$,
(b) $M_{\rm bc}$ distribution for the $\psi'$ region,
(c) $M_{\rm bc}$ distribution for the $X(3823)$ region and
(d) $M_{\rm bc}$ distribution for the $X(3872)$ region. 
The curves used  in the fits are described in~\cite{CURVES}.
}
\end{center}
\end{figure}


\begin{figure}[h!]
\begin{center}
\includegraphics[height=136mm,width=85mm]{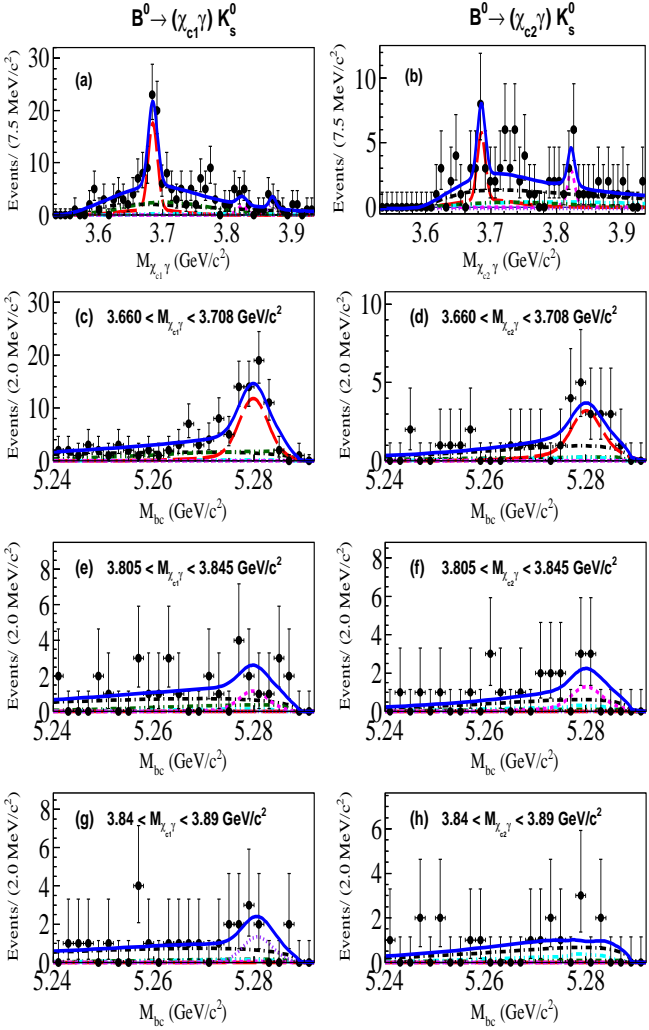}
\caption{\label{fig:cgks}2D UML fit projection 
 for $B^{0} \to (\chicx \gamma) K_S^{0}$ decays:
(a) $M_{\chi_{c1}\gamma}$ distribution for $M_{\rm bc} > 5.27 $ GeV$/c^2$,
(b) $M_{\chi_{c2}\gamma}$ distribution for $M_{\rm bc} > 5.27 $ GeV$/c^2$, 
(c) and (d) $M_{\rm bc}$ distribution for  the $\psi'$ region,
(e) and (f) $M_{\rm bc}$ distribution for  the $X(3823)$ region, and
(g) and (h) $M_{\rm bc}$ distribution for the  $X(3872)$ region.
The curves used in the fits are described in~\cite{CURVES}.}

\end{center}
\end{figure}

\begin{figure}[h!]
\begin{center}
\includegraphics[height=40mm,width=80mm]{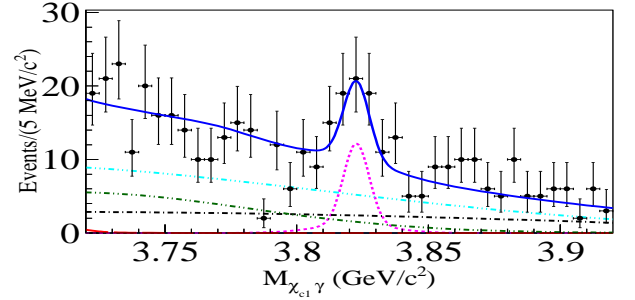}
\caption{\label{fig:sim} 2D UML fit projection of   $M_{\chi_{c1}\gamma}$ 
distribution for the simultaneous fit of  $B^{\pm} \to (\chi_{c1} \gamma) K^{\pm}$ 
and  $B^{0} \to (\chi_{c1} \gamma) K_S^{0}$  decays 
for $M_{\rm bc} > 5.27 $ GeV$/c^2$. 
The curves used in the fits are described in~\cite{CURVES}. }
 
\end{center}
\end{figure}

\begin{table}[h]
\caption{ Summary of the results. Signal yield ($Y$) from the fit, 
significance ($\mathcal{S}$) with systematics included,
corrected efficiency ($\epsilon$) and
measured $\mathcal{B}$.
For $\mathcal{B}$, the first (second) error is statistical (systematic). 
 In the neutral $B$ decay, the efficiency
below includes the $K_S^0 \to \pi^+ \pi^-$ branching fraction but does 
not include the factor of two for $K^0 \to K_S^0~{\rm or}~K_L^0$.}

\begin{center}

  \begin{tabular}{lccccc}
\hline \hline
Decay & Yield ($Y$) & $\mathcal{S} (\sigma)$ & $\epsilon$(\%) & Branching fraction  \\ \hline

\multicolumn{4}{c}{$B^\pm \to \psi'(\to \chicx\gamma)K^\pm$~~~~~~~~~~~~~~~~~} &
$\mathcal{B}(10^{-4})$  \\ \hline
$\chi_{c1}$ & $193.2 \pm 19.2$ & 14.8 & 8.6 & $7.7\pm0.8\pm0.9$  \\
$\chi_{c2}$ & $59.1\pm 8.4 $& 7.8 & 6.0 & $6.3 \pm 0.9 \pm0.6$ \\ \hline

\multicolumn{4}{c}{$B^0 \to \psi'(\to \chicx \gamma)K^0$~~~~~~~~~~~~~~~~~~} &  
 \\ \hline
$\chi_{c1}$ & $50.3 \pm 7.3 $  &  7.2 &  5.1 &  $6.8 \pm 1.0 \pm 0.7$ \\
$\chi_{c2}$ &  $12.9 \pm 4.4 $ &  2.9 & 3.5 & $4.7\pm 1.6\pm0.8$  \\ \hline

\multicolumn{4}{c}{$B^\pm \to X(3823)(\to \chicx \gamma)K^\pm$~~~~~~~~~} & 
$\mathcal{B}(10^{-6})$  \\ \hline
$\chi_{c1}$ & $33.2 \pm  9.7 $ &3.8 & 10.9& $9.7 \pm 2.8 \pm1.1 $ \\
$\chi_{c2}$ & $0.3 \pm 3.9 $ & 0.1 & 8.8 & $<3.6$ \\ \hline

\multicolumn{4}{c}{$B^0 \to X(3823)(\to \chicx \gamma)K^0$~~~~~~~~~~} & 
 \\ \hline
$\chi_{c1}$ & $3.9 \pm 3.4 $ &  1.2 & 6.0&  $<9.9$  \\ 
$\chi_{c2}$ &  $5.3 \pm 2.9 $ & 2.4 & 5.0&  $<22.8$  \\ \hline

\multicolumn{4}{c}{$B^\pm\to X(3872)(\to \chicx \gamma)K^\pm$~~~~~~~~~} & 
& \\ \hline
$\chi_{c1}$ & $-0.9\pm 5.1$ & & 11.1 & $<1.9$  \\
$\chi_{c2}$ & $4.7 \pm 4.4 $ & 1.3  & 9.3 &  $<6.7$ \\ \hline 

\multicolumn{4}{c}{$B^0\to X(3872)(\to \chicx \gamma)K^0$~~~~~~~~~~~} & & \\ \hline
$\chi_{c1}$ &   $4.6 \pm 3.0$ &  1.6 & 6.2 &  $<9.6$  \\
$\chi_{c2}$ &  $2.3 \pm 2.2$ & 1.1 & 5.2 &   $<12.2$ &  \\ \hline \hline

\end{tabular}
\label{tab_results}
\end{center}
\end{table}

A correction for small differences in the signal detection efficiency 
between MC simulation and data has been applied for the lepton  and kaon 
identification requirements. Uncertainties  in these corrections are included 
in the systematic error.  The  $e^+ e^- \to e^+ e^- \ell^+ \ell^-$ 
($\ell =$ $e$ or $\mu$) and
$D^{*+} \to D^0(K^- \pi^+)\pi^+$  samples are used to estimate the lepton 
identification correction and the kaon (pion) identification correction, 
respectively.  To estimate  the correction  and residual systematic uncertainty  
for $K_S^0$ reconstruction, 
 $D^{*+}\to D^0(\to K_S^0 \pi^+ \pi^-) \pi^{+}$ samples are used. The errors on the 
PDF shapes are obtained by varying all fixed parameters by $\pm 1 \sigma$
and taking the change in the yield as the systematic uncertainty. 
The uncertainties 
due to the secondary branching fractions are also taken into account. The 
uncertainties of the tracking efficiency and $N_{B\bar{B}}$ are estimated to be 
0.35$\%$ per track  and $1.4\%$, respectively.  The uncertainty on the photon 
identification is estimated to be 2.0\%$/$photon.
 The systematic uncertainty associated with the
difference  of the $\pi^0$ veto  between data and MC 
is estimated to be 1.2\% from  a study of 
the $B^\pm \to \chi_{c1} (\to J/\psi \gamma) K^{\pm}$ sample.

To improve the mass determination of the $X(3823)$,
a simultaneous fit to $B^{\pm} \to (\chi_{c1}\gamma) K^{\pm}$ 
and
$B^0 \to (\chi_{c1}\gamma) K_S^0$ is performed,   assuming that $\mathcal{B}(B^\pm \to X(3823) K^\pm)/\mathcal{B}(B^0 \to X(3823)K^0)$ = $\mathcal{B}(B^\pm \to \psi' K^\pm)/\mathcal{B}(B^0 \to \psi' K^0)$.
The $\psi'$ peak position and $M_{\chi_{c1}\gamma}$
resolution are common for both charged and neutral $B$ candidates.
 From this fit, we estimate the
significance for $X(3823)$ to be 4.0 $\sigma$  (including systematic
uncertainties).
We determine the mass of the signal peak relative to the well-measured $\psi'$
mass : 
\begin{center}
$M_{X(3823)} = M_{X(3823)}^{\rm meas} - M_{\psi'}^{\rm meas} + M_{\psi'}^{\rm PDG}$  \\
$= 3823.1 \pm 1.8 \pm 0.7$ MeV. \\
\end{center}   
Here, the first uncertainty is statistical and the second is systematic. 
 Because of 
the  mass-constrained fit to the $\chi_{c1}$ candidate, the systematic 
uncertainty of $M_{X(3823)}$ is dominated by the additional photon's energy scale.
This photon energy scale uncertainty is estimated by the difference between
the $\chi_{c1} \to J/\psi \gamma$ candidates' mass without
any constraint and the
$\chi_{c1}$ nominal mass~\cite{pdg}, which results in 0.7 MeV as the $M_{X(3823)}$ 
systematic error. 
In order to 
estimate the width, we float this parameter and  find  no 
sensitivity with the available statistics: the width is $ 1.7 \pm 5.5$ MeV. 
Using pseudo-experiments generated with different width
 hypotheses for the $X(3823)$, the U.L. at 90\% C.L. on  its width
is estimated to be 24 MeV.

%


The mass of  the $X(3823)$ is near potential model expectations
for the centroid of the $1^3D_J$ states: the Cornell~\cite{cornell} 
and the Buchm\"{u}ller-Tye~\cite{buchmuller} potentials give
3810 MeV$/c^2$.  Other models predict the mass of 
$\psi_2$~ (the $1^3D_2~c\bar{c}$ state, having $J^{PC}= 2^{--}$)  to be 
3815-3840 MeV$/c^2$~\cite{Godfrey,Ebert,Eichten,blank}. The $X(3823)$ mass 
agrees quite well with these models. 
In addition, since no peak has been seen around
$X(3823)$ in the $D\bar{D}$ final state~\cite{pakhlov}, 
one expects that $\psi_2$ does not decay to $D \bar{D}$~\cite{Eichten}.
The ratio  
$R_B=\frac{\mathcal{B}(X(3823)\to \chi_{c2}\gamma)}{\mathcal{B}(X(3823)\to \chi_{c1} \gamma)} < 0.41$  (at 90\% C.L.) is consistent with the
expectation ($R_B \sim 0.2$) for $\psi_2$~\cite{pyungwon,qiao,Ebert}. 
 The limited statistics preclude an angular analysis to determine 
the $J^{PC}$ of the $X(3823)$.
The product of branching fractions for  the $X(3823)$ is approximately
two orders of magnitude lower than  for the $\psi'$,  as shown in Table~\ref{tab_results};
it is consistent with the interpretation of  the $X(3823)$ as $\psi_2 (1 ^3D_2)$, whose production  rate is suppressed by the factorization~\cite{fact} in the 
 two-body $B$ meson decays.


In summary, we obtain the first evidence of a narrow state,  $X(3823)$, 
that decays to $\chi_{c1}\gamma$ with  a mass of 
$3823.1\pm1.8({\rm stat})\pm 0.7{\rm(syst)}$ MeV$/c^2$ and  
a significance of 3.8 $\sigma$, including systematic  uncertainties.
We measure the branching fraction product
$\mathcal{B}(B^{\pm} \to X(3823) K^{\pm})  \mathcal{B}(X(3823) \to \chi_{c1}\gamma)$ $=$ $(9.7 \pm 2.8 \pm 1.1)\times 10^{-6}$. No 
evidence is found for $X(3823)\to \chi_{c2}\gamma$ and 
we set an U.L. on its  branching fraction product
$\mathcal{B}$ as well as the ratio 
$R_B\equiv \frac{\mathcal{B}(X(3823)
\to \chi_{c2}\gamma)}{\mathcal{B}(X(3823)\to \chi_{c1}\gamma)} <$ 0.41 
 at 90\% C.L.
 The properties of the $X(3823)$ are consistent with  those
expected for the $\psi_2~(1 ^3D_2~ c\bar{c})$ state. 
We also  determine an U.L. on the product of branching fractions,
${\mathcal B}(B^{\pm} \to X(3872) K^{\pm}){\mathcal B}(X(3872)
\to \chi_{c1} \gamma) < 1.9 \times 10^{-6}$ at 90\% C.L.; this
 is less than one quarter of the corresponding value in 
$X(3872) \to J/\psi \pi^+\pi^-$~\cite{pdg}.
Our results show that the production of the $X(3872)$'s $C$-odd 
partner in two-body $B$ decays and its decay to $\chicx \gamma$ are
considerably suppressed.

We thank the KEKB group for excellent operation of the
accelerator; the KEK cryogenics group for efficient solenoid
operations; and the KEK computer group, the NII, and 
PNNL/EMSL for valuable computing and SINET4 network support.  
We acknowledge support from MEXT, JSPS and Nagoya's TLPRC (Japan);
ARC and DIISR (Australia); FWF (Austria); NSFC (China); MSMT (Czechia);
CZF, DFG, and VS (Germany);
DST (India); INFN (Italy); MEST, NRF, GSDC of KISTI, and WCU (Korea); 
MNiSW and NCN (Poland); MES and RFAAE (Russia); ARRS (Slovenia);
IKERBASQUE and UPV/EHU (Spain); 
SNSF (Switzerland); NSC and MOE (Taiwan); and DOE and NSF (USA).
This work is partly supported by MEXT's Grant-in-Aid for Scientific
Research on Innovative Areas (``Elucidation of New hadrons with a 
Variety of Flavors'').


\begin{thebibliography}{99}
\bibitem{mixchg}
Charge-conjugate and neutral modes are included throughout the paper 
unless stated otherwise.
\bibitem{CX}
Hereinafter, $\chicx$ refers to  either $\chi_{c1}$ or $\chi_{c2}$, depending  on  which is reconstructed.
\bibitem{lg}
The logarithmic Gaussian is parameterized as:\\
$f(x)=\frac{N_0}{\sqrt{2\pi}\sigma_0(\epsilon -x)}
 \exp{ - \frac{(\ln(\frac{\epsilon - x}{\epsilon - x_p}))^2}{2\sigma_0^2} }$,
where $\epsilon = \frac{\sigma}{a} + x_p$ and 
$\sigma_0 =\frac{ \ln(a\sqrt{2\ln 2} + \sqrt{1 + 2a^2\ln 2})}{\sqrt{2\ln 2}}$.  
Here, $N_0$ is the  normalization, $\sigma$ is the standard deviation, 
$x_p$ is the mean and $a$ is the asymmetry.
\bibitem{CURVES}
The curves show the signal [red large-dashed for $\psi'$, magenta dashed for $X(3823)$ and violet dotted for $X(3872)$] and the background component [black dotted-dashed for combinatorial, dark green two dotted-dashed for $B \to\psi' 
(\nrightarrow \chicx \gamma) K $ and cyan three dotted-dashed for peaking component] as well as the overall fit [blue solid]. $B \to\psi' 
(\nrightarrow \chicx \gamma) K $  is specific to the decay mode under study.
\end{thebibliography}

\end{document}